# The Benefits of Spin Polarization for Inertial and Magneto-Inertial Fusion Propulsion

Gerrit Bruhaug [*] [†] and Ayden Kish[‡]
*University of Rochester Laboratory for Laser Energetics, 250 E River Rd., Rochester, NY 14623, USA*

## I. Introduction

Nuclear fusion has long been studied as an energy source for rocketry due to the extremely high energy density and exhaust velocity of the reaction[1–4]. Fusion has four times more specific energy than nuclear fission, which, itself, has two million times more specific energy than chemical fuels. High specific energy allows for rockets that can reach far higher speeds then any chemical rocket - even exceeding the propulsive capability of nuclear fission - while being able to operate with lower mass ratios. This, coupled with the high (approximately 4% of the speed of light) exhaust velocity of fusion-reaction products, puts interstellar travel within the reach of fusion propelled vehicles[2, 5] alongside more near term uses such as interplanetary exploration[3, 4, 7, 8] and planetary defense[6].

Fusion propulsion suffers from two primary complications: the difficulty of igniting a self-sustaining fusion reaction[9–11] and the large amount of ionizing radiation generated by the reaction, which requires a considerable mass of radiation shielding to protect against [1, 9]. Unfortunately, the fuel that is easiest to ignite (deuterium-tritium, or “DT") also produces most of its energy as ionizing radiation in the form of neutrons [1, 9–11]. Fuels that produce fewer neutrons are more difficult to ignite, requiring larger reactors, larger power supplies and more circulating power which adds to system mass[1, 9], potentially outweighing any radiation shielding advantages provided. This paper outlines and derives analytical equations for how a well known nuclear physics technique, spin polarization, can reduce the incident neutron radiation to by a factor of <0.43X of the unpolarized case, reduce fusion ignition requirements (and thus circulating power and power supply mass), increase fusion-gain, and increase the propulsive efficiency of the fusion rocket by 1.01-1.2X depending on nozzle design. The advantages of spin polarization have never been outlined for use in fusion rockets before to the best of the authors’ knowledge. Using spin polarization would allow for lower mass fusion rockets that suffer less radiation damage and heating, while requiring less circulating power and enjoying increased propulsive efficiency.

All nuclei possess an inherent angular momentum known as spin that plays a significant role in nuclear reactions, especially nuclear fusion[12–14]. Spin polarization is the process of aligning the nuclear spin vectors of nuclei. For five-nucleon fusion reactions, notably DT and $D^3He$ (deuterium and helium-3), spin polarization is well known theoretically[12–14] and experimentally[12, 15–17] to increase the fusion cross section by a factor of 1.5X and to force the reaction products to emit anisotropically[12–14]. This cross-section increase is a result of the inherent spins of D and T nuclei adding up to 3/2 in an ideal case. For other four-nucleon fusion reactions, notably DD (deuterium and deuterium)[12], and those with greater than five nucleons, notably $p^{11}B$ (proton and boron-11)[12, 18, 19], there is no direct experimental evidence and quite a contentious theoretical discussions about any potential benefit of spin polarization. Such reactions require more study before any comment can be made on the impact of spin polarization on their use.

[*] Corresponding Author, gbruhaug@ur.rochester.edu
[†] PhD Candidate, Mechanical Engineering Department
[‡] PhD Candidate, Physics and Astronomy Department

Several methods for producing spin-polarized fusion fuel have been considered and tested for macro-scale nuclear physics experiments[22, 23], for the production of beams of spin-polarized particles in particle colliders[12], and for medical applications[30]. These methods range from flowing atomic beams[12] to cryogenic targets 100's of mg in mass[23] to in-situ polarization via polarized laser light[29, 30]. Some consideration has even been given to how to prepare spin polarized targets for inertial fusion experiments[20]. The additional energy consumption from spin polarization is ~100 eV per atom at the most[12, 14], which compares favorably with the >3 MeV per atom from fusion.

Spin polarized fusion research has focused on both magnetic[12, 14] and inertial[21, 24–26] confinement methods and found both methods benefit from spin polarization. It stands to reason that magneto-inertial fusion[3] will also benefit from spin polarization due to the relevant time, field gradient and density scales being within the range of previous spin polarization studies[14, 21, 27, 31, 32]. In spite of the broad range of potential fusion reactor options compatible with spin polarization, this paper will focus entirely on the application of spin polarization to inertial and magneto-inertial fusion propulsion. Purely magnetic confinement systems (such as tokamaks) have much more complex spin depolarization mechanisms at play[14], as well as less-clearly defined gains from the use of spin polarization beyond the increased fusion reactivity due to the complex geometry of the reactor. Magneto-inertial fusion has not been conclusively shown to be compatible with spin polarization and may have higher depolarization rates than expected, but it is reasonable to assume that gains similar to inertial fusion can be had with spin polarization. A case by case analysis of each proposed reactor or rocket would need to be done and that is beyond the scope of this work, however the broad trends from this work should hold for any type of spin polarized fusion rocket.

## II. The Benefits of Spin-Polarized Fusion

As mentioned previously, the well-known favorable response of five-nucleon reactions to spin polarization makes two fuels of particular interest for a spin-polarized fusion rocket: Deuterium-Tritium (DT) and Deuterium-Helium3 ($D^3He$) [12]. Previous beam-target experiments have verified the spin-polarized response of these fusion fuels[15–17] in addition to a robust theoretical understanding[13, 14, 24, 25] of the thermonuclear burn properties. As such, all other fuels will be ignored for the remainder of this paper. The following sections will outline the direct benefits of spin polarization for fusion rockets via newly derived equations using the spatial emission of spin polarized fusion and fits to previous studies on the benefits of spin polarization to fusion reactor performance.

### A. Spin-Polarized DT Benefits

DT fusion is currently the most-promising candidate for near-term use due to its high reactivity and low ignition conditions compared to other fusion fuels [10, 11], and has arguably been the most studied fusion reaction for spin-polarization applications [12–14, 24]. It has also recently been shown to exceed the Lawson criterion in a laboratory setting[35], further motivating the near term focus on DT for fusion rocketry. Fully spin-polarized DT fusion receives a factor of 1.5X increase in fusion cross-section derived from the nuclear physics of the reaction[12–14], which directly translates to increased fuel reactivity. The increase in reactivity provides for either increased fusion energy gain over input energy or decreased input energy for a given fusion gain. Gain increases are estimated to be a factor of ~1.2X up to ~3X for a fixed input energy and are highly dependent on the specifics of the fusion reactor and fuel[21, 24–26]. Conversely the input energy for a given designed gain can be reduced to a factor of ~0.78 to ~0.5 of the original designed input energy. One of the most striking aspects of spin polarized DT fusion is anisotropic neutrons and alpha particle emission[12–14]. Two emission patterns are possible[13], depending on whether the spin vectors of the reacting nuclei are parallel or anti-parallel. For the purposes of this paper, only the parallel emission pattern - as seen in Figure 1 - will be considered, as it is more readily applied to conventional spacecraft designs. This anisotropic emission has been calculated theoretically [13] and seen experimentally [12, 15].

The anisotropic emission has important implications for shielding the spacecraft and for the momentum transfer between the fusion reaction and the spacecraft. This can clearly be seen from the fusion product emission profile compared to the spacecraft shown in Figure 2.

The increased fusion-gain provides a more-economic use of the expensive tritium fuel. The fusion-gain increase factor scales with polarization fraction as shown in Eq. (1) taken from numerical calculations of spin polarized inertial confinement fusion [24]. This factor is multiplied by the gain of a given fusion rocket design to determine the effect of spin polarization on the rocket.

$$I_{\text{Fusion}-\text{Gain}} = 0.45 f_p + 1 \tag{1}$$

Here, $f_p$ is the polarization fraction, ranging from 0 (unpolarized) to 1 (fully polarized). For the reaction of fully polarized DT fusion fuel, the gain is expected to be 145% (factor of 1.45) higher than an equivalent unpolarized case [24]. This directly translates to more 145% more energy from the fuel using the same drivers. Given this increase in specific energy of the reaction for the same amount of fusion fuel, the total specific power of the ship will increase accordingly, increasing total rocket performance. Note that this linear relationship does not hold as the fusion fuel burn-up approaches 100%.

Arguably the largest benefit from spin-polarized DT is the ability to direct the neutron radiation away from the spacecraft. Because 80% of the energy of DT fusion is released in the form of 14.06 MeV neutrons [9, 10], this feature has significant implications for shielding mass requirements and radiator heat loads as can be seen in the comparison between the normalized polarized and unpolarized neutron flux in Figure 3.

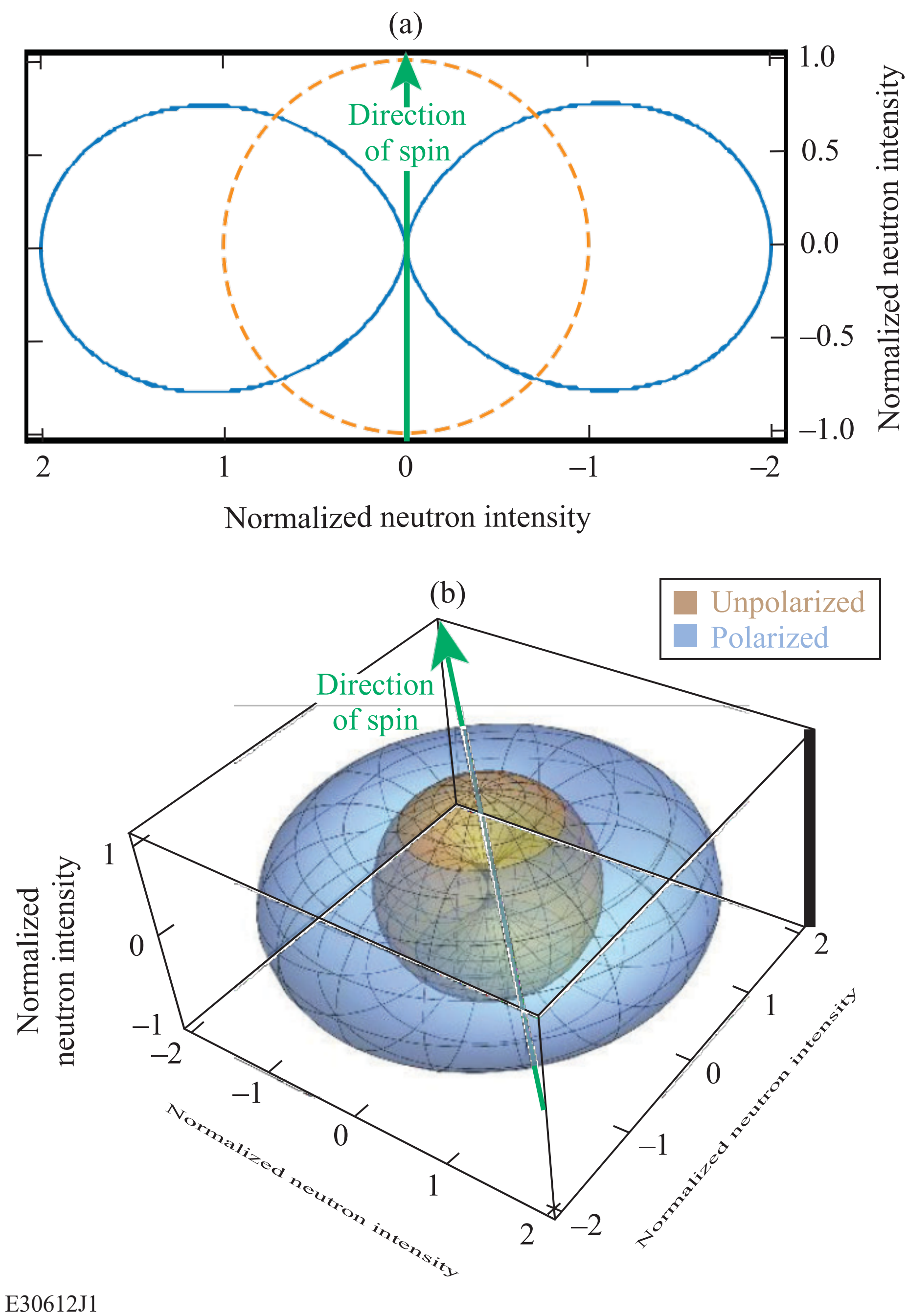

**Fig. 1 Neutron angular emission profile for polarized and unpolarized DT fuel in the center of mass frame.**

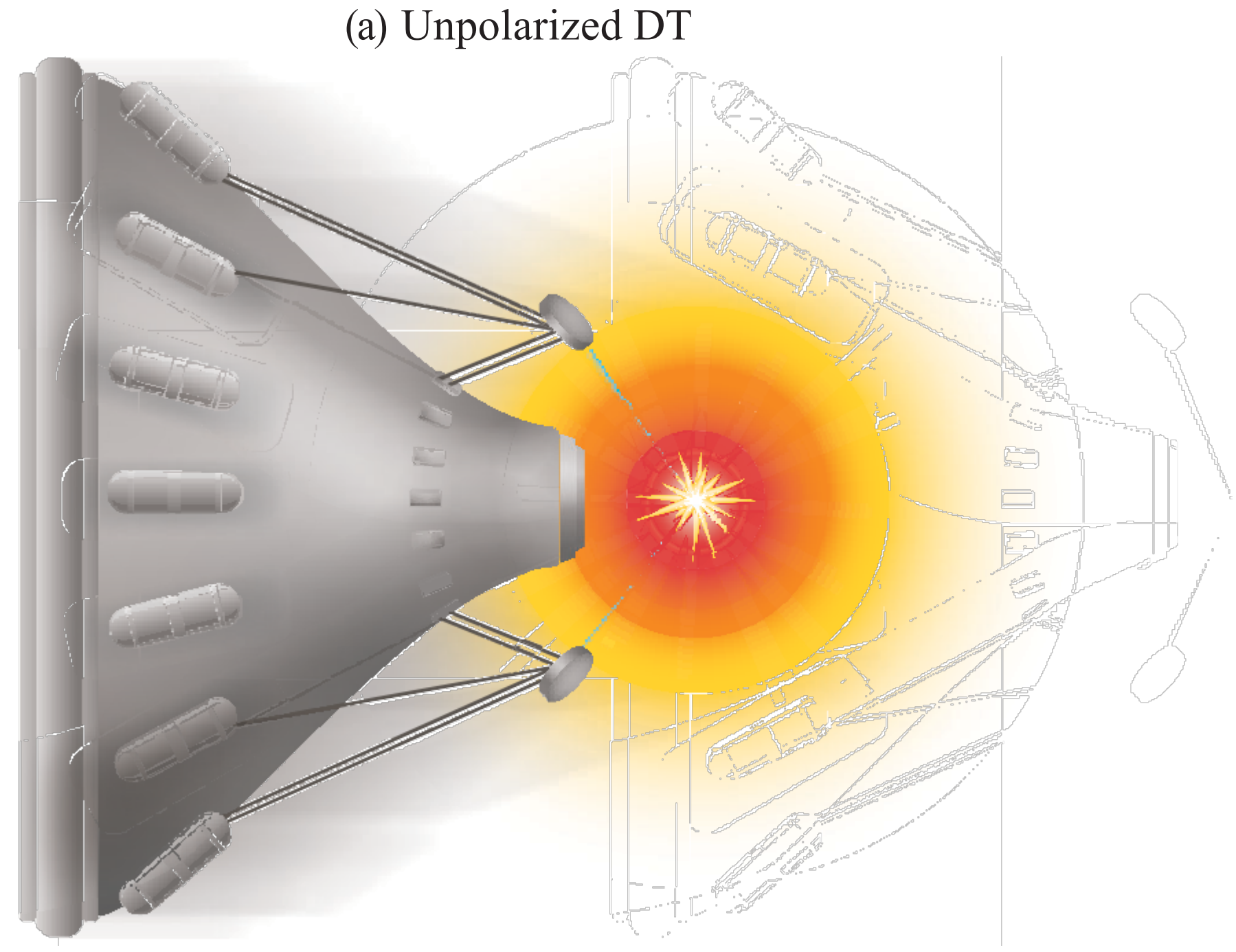

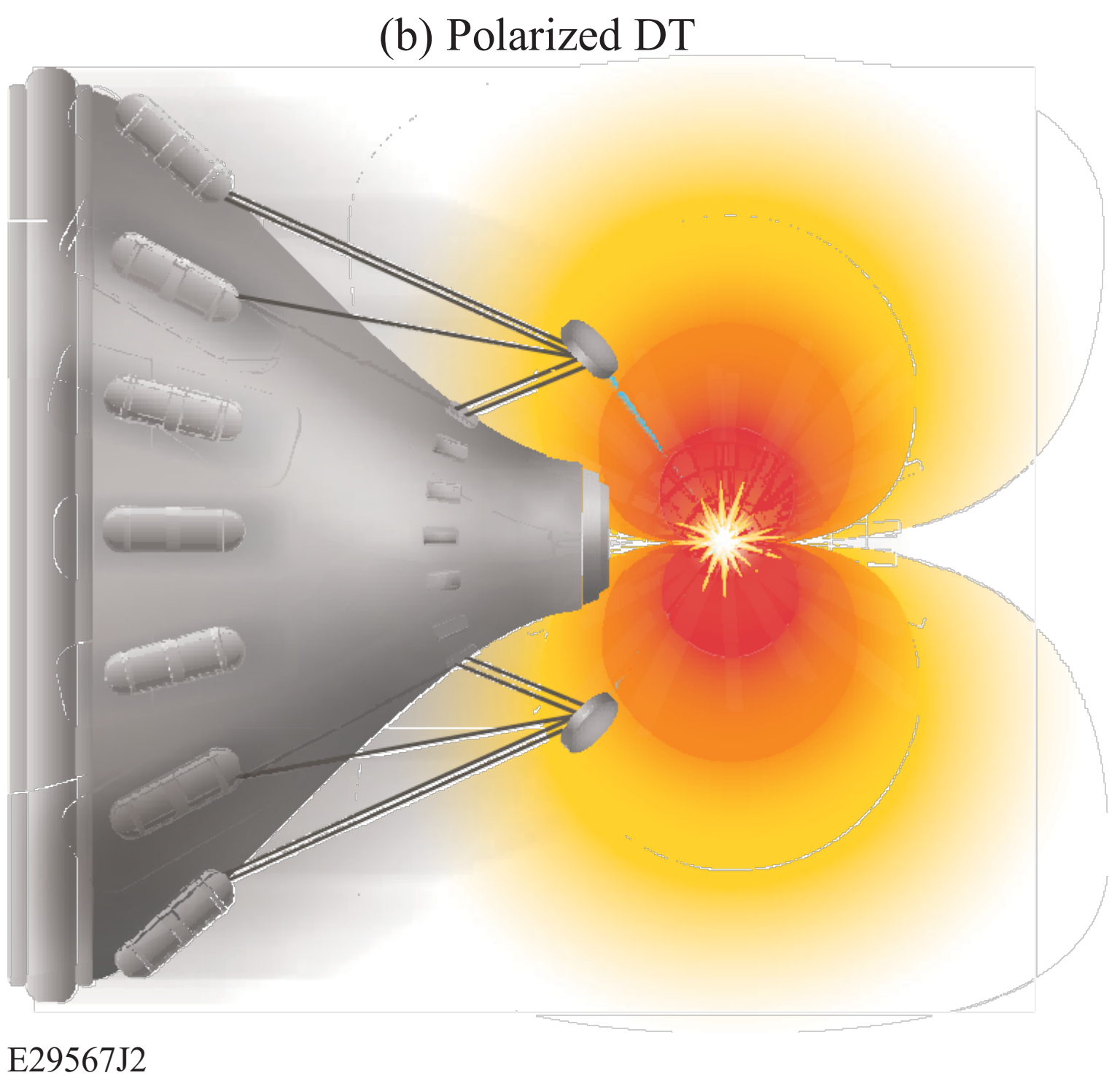

**Fig. 2 A stylized rendition of the neutron and alpha particle emission profile for unpolarized (a) and parallel polarized (b) DT fusion for the VISTA[1] spacecraft.**

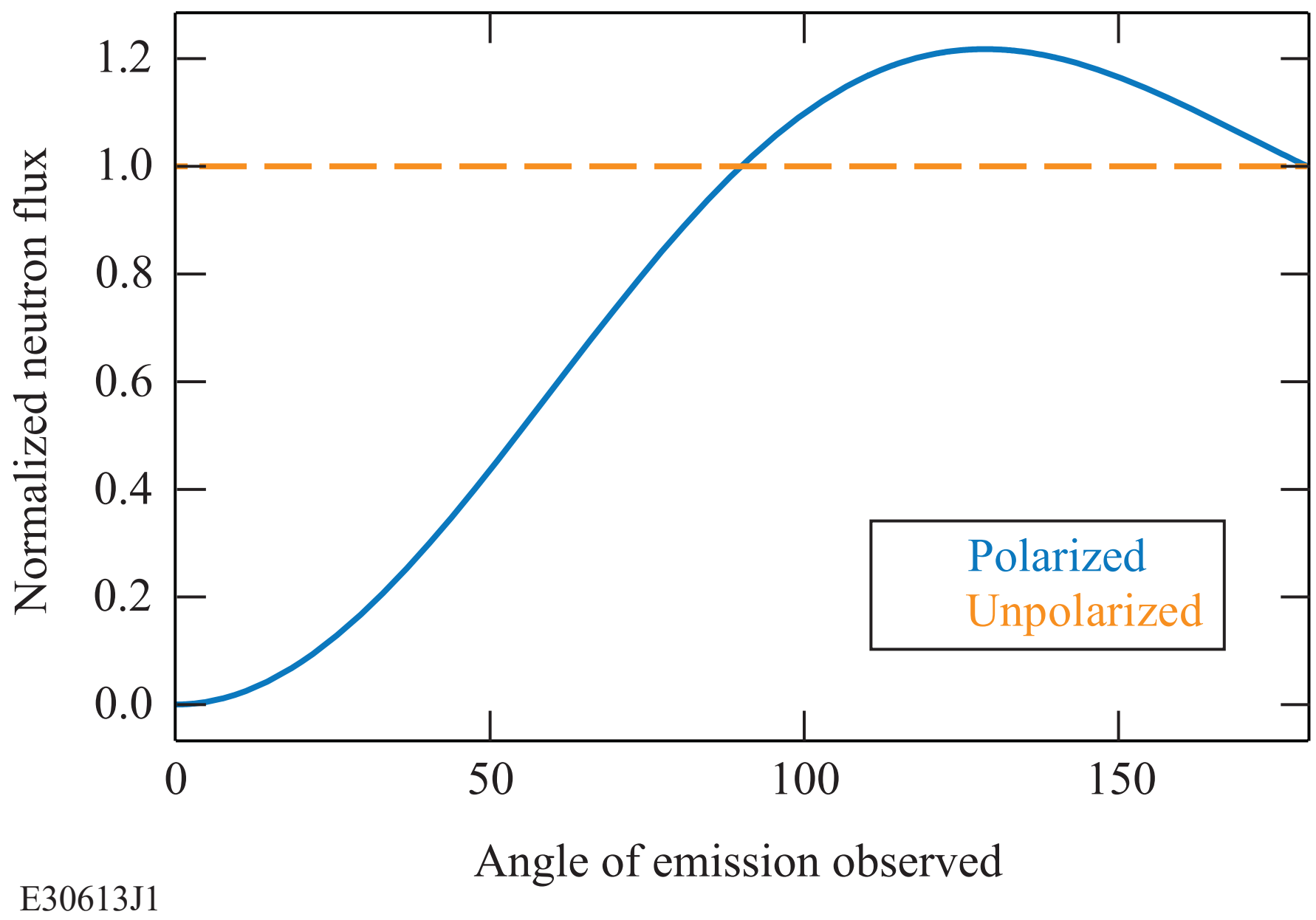

**Fig. 3 Normalized direct neutron emission from DT verses angle of coverage of the ship in polar coordinates for polarized and unpolarized reactions.**

When DT is fully spin polarized, the product neutrons are emitted according to a $\frac{2}{\pi}\sin^2(\theta)$ distribution [12–14]. To calculate the normalized neutron incidence per angle subtended by the spacecraft at fractional polarization one can use Eq (2):

$$E_{\text{NeutronDT}} = \frac{(1 - f_P)}{\theta} \int \frac{1}{\pi} d\theta + \frac{f_P}{\theta} \left( \int \frac{2}{\pi} \sin^2 \theta d\theta \right) = \frac{1}{\pi} \left( 1 - f_P \left( \frac{\cos\theta \sin\theta}{\theta} \right) \right) \tag{2}$$

where the first term corresponds to the unpolarized emission fraction, the second term corresponds to the polarized emission fraction, and $\theta$ is the angle subtended by the ship in radians and $f_P \in [0, 1]$ is the polarization fraction. At full polarization a ship with the angular coverage similar to the VISTA rocket (~50 degrees)[1] has a neutron flux that is a factor of 0.435X reduced compared to the unpolarized case. The VISTA rocket was chosen for this example due to the extensive literature on the concept[1, 5, 28], specifically the well documented shield design, but it is extremely unoptimized for spin polarized fusion. A fusion rocket that subtends ~30 degrees would instead see ~0.17X the neutron flux of the unpolarized case. The reduction in incident neutron seen not only can lower the radiation shielding thickness required but also lowers the resulting neutron heating at a given rocket power that must be rejected by the radiation shield by an equivalent amount. This will result in smaller radiators and higher specific power for a given fusion rocket. The exact savings will vary depending on the specific geometry of the ship in question.

Although there would also be further reductions in neutron emission due to fewer DD fusion side reactions (which do not change neutron emission angles with spin polarization), this is a minor change at best due to the vastly different reactivities of the fuels. On average DT is 100X more reactive than DD, whereas DD also makes 2/3 of the neutrons per reaction that DT does[10, 11]. Thus DD fusion side reaction changes is a sub-percent effect on the neutron load of any spaceship that is not subtending single digit degrees of the neutron emission, and will be ignored for any further analysis of spin polarized DT fusion.

The reduction in shielding mass can be estimated using neutron flux reduction factors (the final flux allowed after the shield divided by the initial flux before it) from other fusion rocket designs[1, 9, 28]. A more detailed analysis would require Monte Carlo simulations[28] that are beyond the scope of this work. To first order the reduction factor for shielding with polarization for DT can be modeled by first normalizing Eq (2) by the unpolarized neutron emission profile to generate Eq (3).

$$N_{\mathrm{DT}} = \left(1 - f_p\left(\frac{\cos\theta\sin\theta}{\theta}\right)\right) \tag{3}$$

This provides an averaged normalized neutron absorption factor for a given ship and polarization fraction. To more conservatively account for the vast difference in neutron flux across the ship, we can then multiply Eq (3) by one minus the neutron distribution and generate Eq (4).

$$N'_{\mathrm{DT}} = \left(1 - f_p\left(\frac{(1-2\sin\theta^2)\cos\theta\sin\theta}{\theta}\right)\right) \tag{4}$$

Then Eq (4) can be be used in to determine the shielding reduction factor via Eq (5).

$$R_{\mathrm{Shielding,DT}} = \frac{log[\frac{R_f}{N'_{DT}}]}{log[R_f]} \tag{5}$$

Here the flux reduction factor is $R_f$, and is taken to be the factor that the neutron flux must be reduced for material degradation or biologic safety reasons. Using the VISTA fusion rocket flux reduction factor[1, 28] of $8 \times 10^{-5}$ the neutron shielding thickness can be increased by a factor of 1.01X or reduced to <0.87X for subtended angles from 50 degrees to 20 degrees. It can be seen that subtending as large of an angle as VISTA will provide no shielding thickness reduction using this conservative analysis, but a spacecraft that subtended a smaller angle could see significant reduction in shielding mass. For the thickness's of LiH and Li used in the VISTA design[1, 28] the smallest angles would save $> 140\ kg/m^2$ and $> 90\ kg/m^2$ respectively. An in depth analysis of each rocket design will be needed to determine the exact shielding savings from spin polarization due to differences in magnetic coil damage assumptions[1, 2, 8, 9]. All fusion reactions also produce gamma rays and x-rays [1, 10, 26, 28] that are unaffected by the spin polarization and will still need to be stopped by the same thickness of shielding no matter if the fuel is polarized. However, the reduction in neutron irradiation will vastly lower the waste heat requirements for the ship[1] and thus lower the low temperature radiator mass requirements.

The other potential benefit from spin polarization of fusion fuel is the change in momentum coupling between the fusion fuel and the spacecraft at high burn-up fractions. In a typical pulsed fusion rocket the charged alpha particles explode outwards isotropically, heating the remaining fuel and inert mass and causing all of the reaction mass to expand outwards equally in all directions. A portion of this expanding plasma then is redirected by a magnetic field to produce thrust for the spacecraft [1, 2, 4, 9]. In the case of spin polarized DT fusion the alpha particles will be emitted in the opposite direction of each neutron to conserve momentum, and thus with the same angular distribution seen in Fig 1. [12, 13]. At extremely high fusion burn-ups, this will lower the losses in the magnetic loss cone of a typical magnetic nozzle of a fusion rocket [1, 4, 9] and increase the total propulsive efficiency of the nozzle as fewer fusion products are directed towards the lossy central portion of the magnetic nozzle. Further propulsive gains may be had with a more complex magnetic nozzle, an electrostatic nozzle [34], or even a magnetic sail [33] designed to better utilize the unique emission profile from spin polarized fusion.

To model the nozzle efficiency enhancement we will simply look at the change in alpha particles (and any plasma they push with their expulsion) lost to the magnetic nozzle loss cone. The alpha emission will mirror the neutron emission shown in Figure 1[12, 13] due to conservation of momentum, so the same analysis technique used for Eq (2) and (3) can be used. We then compare the change in lost alpha particles to the unpolarized case and derive Eq (6) for the change in losses into the nozzle loss cone. The relative propulsive gains can then be modeled via Eq (7) where the change in losses is normalized by the unpolarized case.

$$f_{loss} = \left(1 - f_p\right) \int \frac{1}{\pi} d\theta + f_p \left( \int \frac{2}{\pi} \sin^2 \theta d\theta \right) = \frac{\theta - f_p \cos\theta \sin\theta}{\pi} \tag{6}$$

$$I_{\text{Nozzle}} = \frac{1 - f_{loss,pol}}{1 - f_{loss,unpol}} = \left[ 1 - \frac{\theta - f_p \cos\theta \sin\theta}{\pi} \right] / \left[ 1 - \frac{\theta}{\pi} \right] \tag{7}$$

This equation provides the relative gain factor of propulsive efficiency (i.e. both thrust and specific impulse) of any externally ignited fusion rocket scheme as a function of polarization and magnetic loss cone angle. Here $\theta$ is the angle of the loss cone and $f_p$ is polarization fraction. The results of this equation were plotted in Figure 4 for various magnetic loss cone angles. Depending on the loss cone angles the propulsive efficiency can increase by 1.01X to >1.2X of the unpolarized case.

**B. Spin Polarized $D^3He$ Benefits**

$D^3He$ fusion is the fusion reaction most commonly invoked when one wants to avoid neutron radiation[2, 8, 9, 25]. Although the primary reaction does not generate neutrons, DD side reactions will generate neutrons that will still irradiate the ship [1, 9, 10]. As $D^3He$ is not much more reactive than DD (5-10× depending on temperature), compared to DT's 100× [10, 11] higher reactivity, the number of DD side reactions will be far higher in a burning $D^3He$ plasma versus a burning DT plasma. The ratio of D to $D^3He$ can be lowered to lower neutron production[8], but this reduces fusion reactivity and increases x-ray losses. The higher ignition requirements of $D^3He$ will also typically require a larger driver and/or larger fusion reactor [9, 25]. This can translate to larger shielding masses due to more surface area needing to be shielded on top of the higher mass drivers and associated radiators [9]. Still, spin polarization offers the same benefit of a 1.5X increase in cross section to $D^3He$ fusion as it does to DT fusion [12, 25], as well as the anisotropic primary particle emission. The same increases in gain or reductions in input energy also hold for $D^3He$[25], which is arguably the most compelling reason to utilize spin polarization in $D^3He$ fusion rockets.

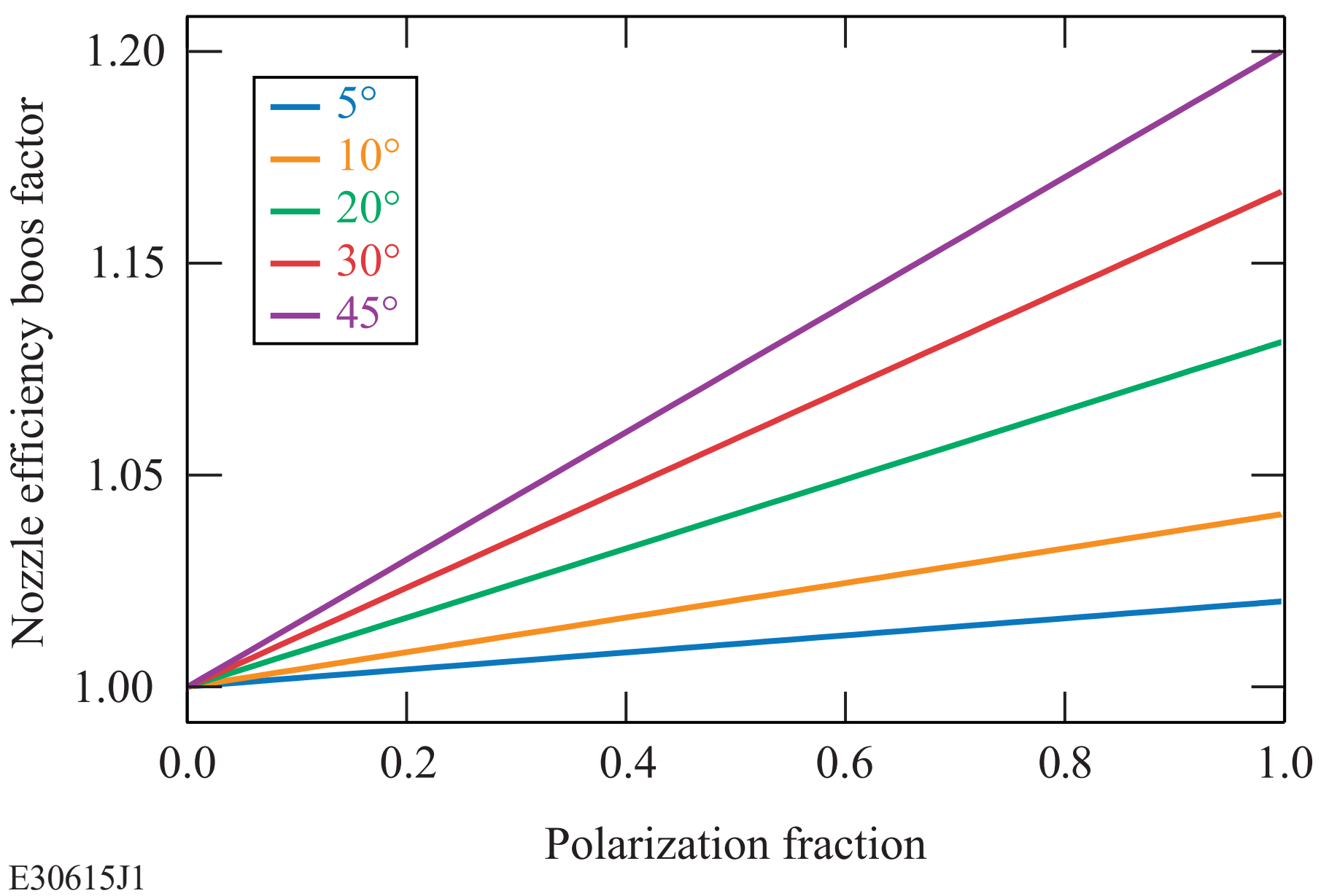

**Fig. 4 Propulsive gain factor verses polarization fraction for various magnetic nozzle loss cone angles**

The same approach used for DT radiation shielding reduction will be used for $D^3He$, but the only reduction in radiation dose comes from the reduction in DD secondary reactions due to the increased burn-up of the primary fuel. The normalized neutron reduction versus polarization is shown in Eq. (8) and is unchanged by the angle subtended by the spacecraft. The same shielding logic as was used for DT can then be used to derive the shielding reduction factor with polarization fraction for $D^3He$ shown in Eq. (9).

$$N_{\mathrm{D^3He}} = 1 - 0.2424 f_p \tag{8}$$

$$R_{\mathrm{shielding,D^3He}} = \frac{log[\frac{R_f}{N_{D^3He}}]}{log[R_f]} \tag{9}$$

Note that the lack of dependence on the angle subtended by the ship is due to the fact that the neutrons are only produced by secondary and tertiary reactions, which will not be spin polarized. At full polarization, the neutron load of the ship is reduced to 0.757X of the unpolarized case, which reduces the neutron related heating by and equivalent amount, as seen by Eq (5). This translates to a reduction in shielding thickness of 3%, or a factor of 0.97X of the original thickness, which is not as great of a saving as DT, but the total shielding thickness initially needed is also lower[2, 8, 9]. The common assumption of a $D^3He$ reactor being physically larger than a DT reactor due to the higher ignition requirements [9–11, 25] also may end up providing more total shielding mass reductions to $D^3He$ fueled rockets than DT fueled rockets depending on reactor design. A comparison of neutron shielding thickness reduction factors between DT for various subtended angles and $D^3He$ can be seen in Figure 5.

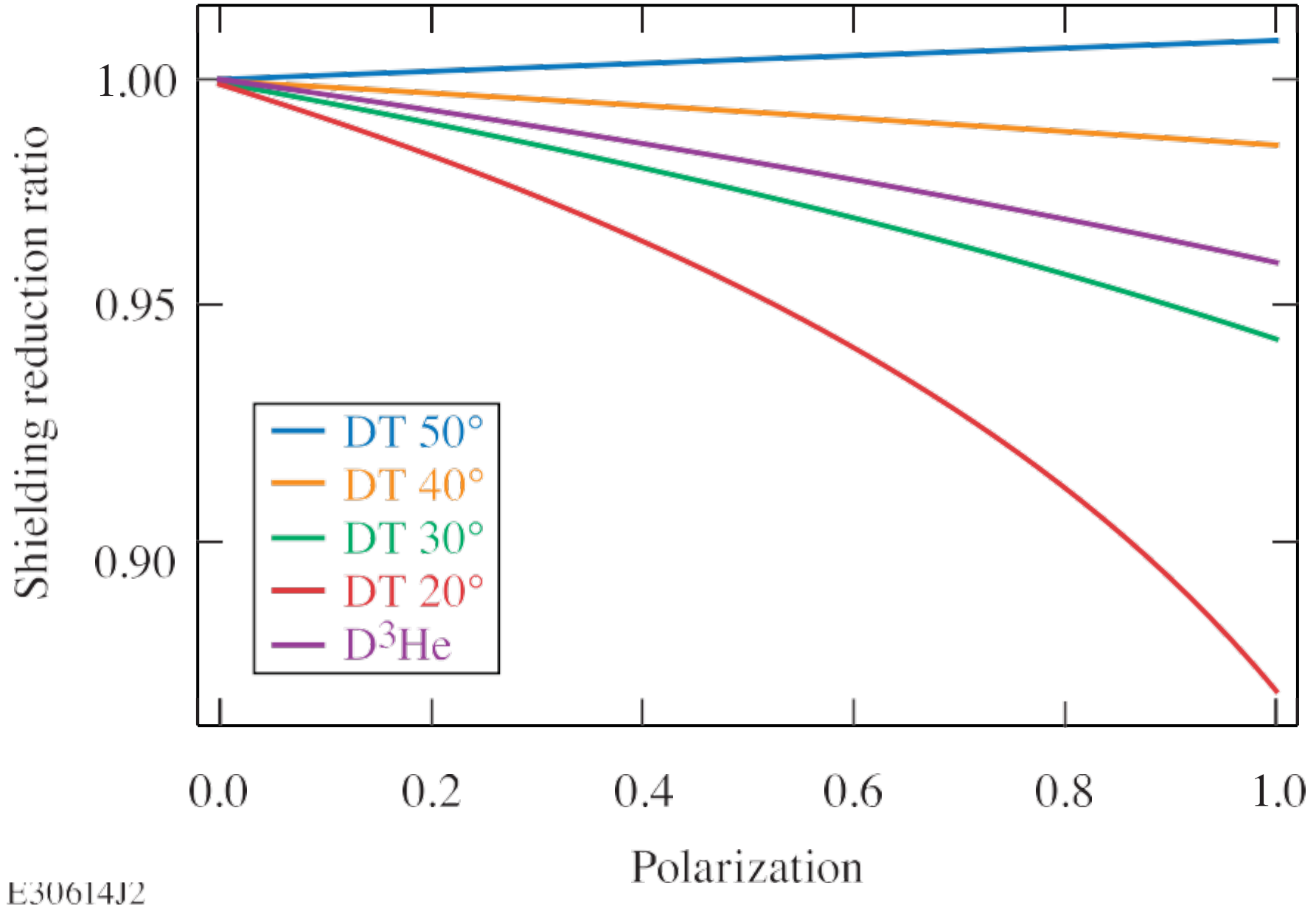

**Fig. 5 Comparison of shielding reduction factor from spin polarization for DT and $D^3He$ fusion rockets using VISTA neutron flux reduction factors and angular coverage**

It can be seen that DT fusion enjoys a greater shielding reduction factors than $D^3He$ for a given polarization fraction if the ship subtends an angle ≤ 30. How this effects final fusion rocket mass will depend on rocket design specifics because $D^3He$ rockets require less initial shielding thickness to begin with, but may be overall larger due to the larger ignition requirements [1, 2, 9, 25].

$D^3He$ fusion will also benefit from the same increase in propulsive efficiency as was outlined above for DT in Eq. (7), and the equations will remain the same. The various rocket enhancements (lower ignition requirements or lower driver circulating power and increased propulsive efficiency) are arguably a much larger motivation for using spin-polarized $D^3He$ fuel rather than the reduction in neutron radiation, while for DT fuel the reduction in neutron radiation is more compelling due to the high fraction of energy released as neutron radiation[1, 9, 10].

## III. Conclusion

As outlined in this paper, spin-polarized fusion has the potential to provide a reduction in the neutron shield thickness (potentially by a factor of 0.87X or better) of fusion rockets while also lowering radiation damage and heating of the ship.

The use of spin polarization also provides more fusion gain and/or a reduction in required input energy [21, 24–26]. Spin-polarization directs fewer fusion exhaust products into the loss cone of magnet nozzles, increasing propulsive efficiency by a factor of 1.01-1.2X. Recent advancements in spin polarization for nuclear physics experiments have shown that the technology is rapidly maturing [12, 23, 29, 30] and has potential to be implemented on near-term fusion rockets. The advantages of spin polarized fusion are primarily seen by DT fueled fusion rockets due to the anisotropic neutron emission, but rockets utilizing $D^3He$ may also choose to use spin-polarized fuel to lower the prodigious ignition requirements of this advanced fuel [9, 10, 25] while also increasing propulsion efficiency and lowering radiation shielding thickness. Spin-polarized fusion can be seen to be a straight forward method of increasing the performance of many fusion rocket designs and should be considered when designing future spacecraft.

## Funding Sources

This research did not receive any specific grant from funding agencies in the public, commercial, or not-for-profit sectors. It was done entirely as a volunteer project by the authors.

## Acknowledgments

The lead author thanks Tanner Horne for introducing him to the concept of spin polarized fusion and to the wonderful online community at ToughSF for stimulating discussions about fusion propulsion that led to this paper. The authors would also like to personally thank Jeffrey K. Greason for enlightening discussions and for pointing out the potential boost in propulsive capability of spin polarized fusion and Andrew J. Higgins for pointing out crucial and longstanding errors in the community understanding of spin polarization geometry. Lucas Beveridge also did a great service in reviewing many versions of this manuscript.

## References

[1] Orth, C. D., "VISTA – A Vehicle for Interplanetary Space Transport Application Powered by Inertial Confinement Fusion," Tech. Rep. UCRL–TR–110500, Lawrence Livermore National Laboratory, Livermore, CA, May 2003.

[2] Long, K. F., Obousy, R. K., and Hein, A., "Project Icarus: Optimisation of Nuclear Fusion Propulsion for Interstellar Missions," Acta Astronautica, Vol. 68, 2011, pp. 1820–1829. https://doi.org/10.1016/j.actaastro.2011.01.010.

[3] Miernik, J., Statham, G., Fabisinski, L., Maples, C. D., Adams, R., Polsgrove, T., Fincher, S., Cassibry, J., Cortez, R., Turner, M., and Percy, T., "Z-Pinch Fusion-Based Nuclear Propulsion," Acta Astronautica, Vol. 82, No. 11-12, 2013, pp. 173–182. https://doi.org/10.1016/j.actaastro.2012. 02.012.

[4] Cassenti, B., Budica, R., Johnson, L., and Kammash, T., "From Laser Pulse Propulsion to Fusion Pulse Propulsion: An Evolutionary Approach," 51rst Joint Propulsion Conference and Exhibit, AIAA Paper 2015-3858, July 2015, pp.1–9. https://doi.org/10.2514/6.2015-3858.

[5] Long, K. F., “Sunvoyager: Interstellar Precursor Probe Mission Concept Driven by Inertial Confinement Fusion Propulsion,” Journal of Spacecraft and Rockets, Vol. 0, No. 0, 0, pp. 1–15. https://doi.org/10.2514/1.A35539, URL https://doi.org/10.2514/1.A35539.

[6] Wurden, G. A., Weber, T. E., Turchi, P. J., Parks, P. B., Evans, T. E., Cohen, S. A., Cassibry, J. T., and Campbell, E. M., “A New Vision for Fusion Energy Research: Fusion Rocket Engines for Planetary Defense,” Journal of Fusion Energy, Vol. 35, No. 5, 2016, pp. 123–133. https://doi.org/10.1007/s10894-015-0034-1.

[7] Cassibry, J., Cortez, R., Stanic, M., Watts, A., Seidler, W. I., Adams, R., Statham, G., and Fabisinki, L., “Case and Development Path for Fusion Propulsion,” Journal of Spacecraft and Rockets, Vol. 52, No. 2, 2015. https://doi.org/10.2514/1.A32782.

[8] Razin, Y. S., Pajer, G., Breton, M., Ham, E., Mueller, J., Paluszek, M., Glasser, A. H., and Cohen, S. A., “A direct fusion drive for rocket propulsion,” Acta Astronautica, Vol.105, No. 1, 2014, p. 145–155. https://doi.org/10.1016/j.actaastro.2014.08.008.

[9] Haloulakos, V., and Bourque, R. F., “Fusion Propulsion Study,” Tech. Rep. ADA212935, McDonnell Douglas Space Systems Co., Huntington Beach, CA, July 1989.

[10] Atzeni, S., and Meyer-Ter-Vehn, J., “The Physics of Inertial Fusion: Beam Plasma Interaction, Hydrodynamics, Hot Dense Matter,” International Series of Monographs on Physics, Vol. 125, Oxford University Press, Oxford, 2004, 1st ed., pp. 2–25. https://doi.org/10.1103/PhysRevA.48. 4461.

[11] Bosch, H., and Hale, G., “Improved Formulas for Fusion Cross-Sections and Thermal Reactivities,” Nuclear Fusion, Vol. 32, April 1992, pp. 611–631. https://doi.org/10.1088/0029-5515/32/4/I07.

[12] Ciullo, G., Engels, R., Büscher, M., and Vasilvev, A., "Nuclear Fusion with Polarized Fuel", Vol. 187, Springer International Publishing, Switzerland, July 2016, https://doi.org/10.1007/978-3-319- 39471-8.

[13] Hupin, G., Quaglioni, S., and Navrátil, P., “Ab Initio Predictions for Polarized Deuterium-Tritium Thermonuclear Fusion,” Nature Communications, Vol. 10, No. 1, Dec. 2019, p. 351. https://doi.org/10.1038/s41467-018-08052-6.

[14] Kulsrud, M., Valeo, E., and Cowley, S., “Physics of Spin- Polarized Plasmas,” Nuclear Fusion, Vol. 26, No. 11, 1987, pp. 1443–1462. https://doi.org/10.1088/0029-5515/26/11/001.

[15] Bem, P., Kroha, V., Mares, J., Simeckova, E., Trginova, M. and Vercimak, P., "Angular Anisotropy of the 3H(d, alpha) n Reaction at Deuteron Energies Below 200 keV," Few-Body Systems 22, 77–89 (1997). https://doi.org/10.1007/s006010050055

[16] Geist, W.H., Brune, C.R., Karwowski, H.J, Ludwig, E.J., Veal, K.D., "The 3He(d,p)4He reaction at low energies", Phys. Rev. C, Vol. 60, 1999, p. 054003, https://doi.org/10.1103/PhysRevC.60.054003

[17] Braizinha, B., Brune, C.R., Fisher, B.M., Karwowski, Leonard, D.S., Ludwig, E.J., Santos, F.D., Thompson, I.J., "Resonant and direct components in the 3He(d,p)4He reaction at low energies", Phys. Rev. C, Vol. 69, 2004, p. 024608, https://doi.org/10.1103/PhysRevC.69.024608

[18] Dmitriev, V.F., "Effect of Polarization on the Cross Section for the Reaction 11B(p,alpha)8Be and Angular Distributions of Its Products", Physics of Atomic Nuclei, 69(9), 2006, 1461-1462. doi:101134/S1063778806090043

[19] Ahmed, M.W., Weller, H.R., "Nuclear Spin-Polarized Proton and 11B Fuel for Fusion Reactors: Advantages of Double Polarization in the 11B(p, alpha)8Be Fusion Reaction", Journal of Fusion Energy, 33(2), 103–., 2014, https://doi.org/10.1007/s10894-013-9643-8

[20] Goel, B., Herringa, W., "Spin Polarized ICF Targets", Nucl. Fusion 28 355, 1998, https://doi.org/10.1088/0029-5515/28/3/001

[21] Pan, Y., and Hatchett, S. P., "Spin-Polarized Fuel in High Gain ICF Targets," Nuclear Fusion, Vol. 27, No. 5, 1987, pp. 815–819. https://doi.org/10.1088/0029-5515/27/5/010.

[22] Shinkoda, I., Reynolds, M. W., Cline, R. W., and Hardy, W. N., "Measurements on Mixtures of Doubly Spin-Polarized Hydrogen and Doubly Spin-Polarized Deuterium," Japanese Journal of Applied Physics, Vol. 26, No. 3-1, 1987, pp. 243–244. https://doi.org/10.7567/JJAPS.26S3.243

[23] Bass, C., Bade, C., Blecher, M., Caracappa, A., D'Angelo, A., Deur, A., Dezern, G., Glueckler, H., Hanretty, C., Ho, D., Honig, A., Kageya, T., Khandaker, M., Laine, V., Lincoln, F., Lowry, M., Mahon, J., O'Connell, T., Pap, M., Peng, P., Preedom, B., Sandorfi, A., Seyfarth, H., Stroeher, H., Thorn, C., Wei, X., and Whisnant, C., "A portable cryostat for the cold transfer of polarized solid HD targets: HDice-I," Nuclear Instruments and Methods in Physics Research Section A: Accelerators, Spectrometers, Detectors and Associated Equipment, Vol. 737, 2014, pp. 107–116. https://doi.org/https://doi.org/10.1016/j.nima.2013.10.056.

[24] Temporal, M., Brandon, V., Canaud, B., Didelez, J. P., Fedosejevs, R., and Ramis, R., "Ignition Conditions for Inertial Confinement Fusion Targets with a Nuclear Spin-Polarized DT Fuel," Nuclear Fusion,Vol. 52,No. 10, 2012, p.103011. https://doi.org/10.1088/0029-5515/52/10/103011.

[25] Honda, T., Nakao, Y., Honda, Y., Kudo, K., and Nakashima, H., "Burn Characteristics of Inertially Confined D-3He Fuel," Nuclear Fusion, Vol. 31, No. 5, 1991, pp. 851–865. https://doi.org/10.1088/0029-5515/31/5/004.

[26] More, R. M., "Nuclear Spin-Polarized Fuel in Inertial Fusion," Physical Review Letters, Vol. 51, No. 5, 1983, pp. 396–399. https://doi.org/10.1103/PhysRevLett.51.396.

[27] Ronghao, H., Hao, Z., Zhihao, T., Zhihao, Z., Meng, L., Shiyang, Z, Ding, Yongkun, D., "Numerical Study of Spin-Polarized Deuterium-Tritium Fuel Persistence in Inertial Confinement Fusion Implosions", Physical Review Research, Vol. 5, 2023, p. 033115, https://doi.org/10.1103/PhysRevResearch.5.033115.

[28] Şahin, S., and Mehmet Şahin, H., "Optimization of the Radiation Shielding Mass for the Magnet Coils of the VISTA Spacecraft," Annals of Nuclear Energy, Vol. 28, No. 14, 2001, pp. 1413–1429. https://doi.org/10.1016/ S0306-4549(00)00134-1.

[29] Spiliotis, A., Xygkis, M., Koutrakis, M., Boulogiannis, G. K., Chrysovalantis, S. K., Katsoprinakis, G. E., Dimitrios, S., and Rakitzis, T. P., "Ultrahigh-density spinpolarized hydrogen isotopes from the photodissociation of hydrogen halides: new applications for laser-ion acceleration, magnetometry, and polarized nuclear fusion," Light: Science and Applications, Vol. 10, No. 35, 2021. https://doi.org/10.1038/s41377-021-00476-y.

[30] Gentile, T., Nacher, P. J., Saam, B., and Walker, T. G., "Optically polarized 3He," Review of Modern Physics, Vol. 89, No. 4, 2017, p. 045004. https://doi.org/10.1103/RevModPhys.89.045004.

[31] Shearer, L., and Walters, G., "Nuclear spin-lattice relaxation in the presence of magnetic-field gradients," Physical Review, Vol. 139, 1965, pp. A1398–A1402. https://doi.org/https://doi.org/0.1103/PhysRev.139.A1398.

[32] Ronghao, H., Zhou, H., Zhihao, T., and Lv, M., "Spin depolarization induced by self generated magnetic fields during cylndrical implosions," Physical Review E, Vol. 102, 2020, p. 043215. https://doi.org/10.1103/PhysRevE.102. 043215, URL https://link.aps.org/doi/10.1103/PhysRevE.102.043215.

[33] Andrews, D., and Zubrin, R., "Nuclear device-pushed magnetic sails (MagOrion)," 33rd Joint Propulsion Conference and Exhibit, 1997, p. 97. https://doi.org/10.2514/6.1997-3072, URL https://arc.aiaa.org/doi/abs/10.2514/6.1997-3072.

[34] Jackson, G. P., "Antimatter-Based Propulsion for Exoplanet Exploration," Nuclear Technology, Vol. 208, No.sup1, 2022, pp. S107–S112. https://doi.org/10.1080/00295450.2021.1997057, URL https://doi.org/10.1080/00295450.2021.1997057.

[35] Abu-Shawareb, H., Acree, R., Adams, P., Adams, J., Addis, B., Aden, R., Adrian, P., Afeyan, B. B., Aggleton, M., Aghaian, L., Aguirre, A., Aikens, D., et al., "Lawson Criterion for Ignition Exceeded in an Inertial Fusion Experiment," Phys. Rev. Lett., Vol. 129, 2022, p. 075001. https://doi.org/10.1103/PhysRevLett.129.075001, URL https://link.aps.org/doi/10.1103/PhysRevLett.129.075001.